\documentclass[preprint,3p,12pt]{elsarticle}

\usepackage{amssymb}
\usepackage{amsmath}
\usepackage{url}
\usepackage{subcaption} 
\usepackage{caption}

\journal{Journal of Computational Physics}

\begin{document}

\begin{frontmatter}



\title{A Concurrent Multiscale Framework Coupling Direct Simulation Monte Carlo and Molecular Dynamics\tnoteref{RR}}
\tnotetext[RR]{LLNL-JRNL-2006036}

\author[UCD,LLNL]{Tim Linke} 
\ead{talinke@ucdavis.edu}
\author[LLNL]{Dane Sterbentz}
\author[UCD]{Niels Grønbech-Jensen}
\author[UCD]{Jean-Pierre Delplanque}
\author[LLNL]{Jonathan Belof}

\affiliation[UCD]{organization={University of California, Davis},
            addressline={1 Shields Ave}, 
            city={Davis},
            postcode={95616}, 
            state={California},
            country={USA}}
\affiliation[LLNL]{organization={Lawrence Livermore National Laboratory},
            addressline={7000 East Ave}, 
            city={Livermore},
            postcode={94550}, 
            state={California},
            country={USA}}

\begin{abstract}
We present a new method to couple the Direct Simulation Monte Carlo (DSMC) algorithm with molecular dynamics (MD). The coupling approach generalizes prior coupling methods using a cell-based decision. The approach is supported by a lifting and restricting operator, which translate length and time scales between DSMC and MD. We verify the framework on basic conservation laws, and demonstrate its usability on a hypersonic flow example. Its accuracy gain is discussed in light of conventional DSMC simulations. Advantages and limitations are discussed, and additional use cases are highlighted.
\end{abstract}




\end{frontmatter}



\section{Introduction}
When modeling a fluid flow field, an essential decision to make is how to treat deviations from equilibrium. Continuum mechanics is widely adopted in computational fluid dynamics, but many cases exist where large gradients occur in very short length and time scales, challenging the continuum assumption. A meaningful measure for the significance of these effects is captured in the Knudsen number; the ratio of the mean free path of a flow's particle to a characteristic length. High Knudsen number flows are of great interest in fields such as reentry and low-orbit aerodynamics\cite{reentry}, microfluidics\cite{microflows}, plasmas\cite{plasmas}, and porous materials\cite{porous}. The validity of continuum mechanics at these conditions is found in the formulation of fluxes, which are assumed to be proportional to gradients of velocity and temperature. However, in high Knudsen number flows (Kn$>$0.1), particle collisions occur infrequently over an observed distance and thus restrict the achievement of local equilibrium. Velocity and temperature gradients therefore contain significant fluctuations that are not captured in continuum mechanics. \cite{Gallis}

To fully capture the non-continuum behavior, a microscopic analysis must replace the gradient-based approach. From a molecular perspective, deviations from equilibrium can be captured using velocity distribution functions $f(x,v,t)$, where $x$ denotes the position and $v$ the velocity of particles at time $t$. The Boltzmann equation describes the evolution of these distributions $f$ with
\begin{equation} \label{eq:Boltzmann}
\frac{\partial f}{\partial t} + v \frac{\partial f}{\partial x} + \frac{F}{m} \frac{\partial f}{\partial v} = (\frac{\delta f}{\delta t})_c ,
\end{equation}
where $F$ is an external force, and the right hand side is the collision operator. The differentials of $f$ describe the change in phase space volume due to particle motion that is uncorrelated with collisions. Molecular interactions are described by the collision operator. The Boltzmann equation is established as the main description of moderate to high Knudsen number flows, and stands as a pillar to the kinetic theory of gases and plasmas \cite{Lee2019}\cite{Sone}.

While its original formulation was designed for simple gases with short-range interactions, adjustments can be added for more complex behavior. These include evaporation and condensation effects\cite{Sone2}, chemical reactions and the inclusion of thermal radiation\cite{Bird1994}. Unfortunately, the nonlinearity of these adjustments make the analytical solution of the Boltzmann equation infeasible for practical applications \cite{Kardar}.

In response, various numerical methods have been developed to solve the Boltzmann equation\cite{Sone}. The moment method involves expanding the distribution function in terms of its moments, such as density, momentum, and energy, which results in a hierarchy of coupled partial differential equations for these macroscopic quantities that can be truncated and solved\cite{moment}. The method excels near equilibrium conditions, but cannot state the accuracy of the approximated solution. In turn, the Knudsen number expansion method readily offers a solution at arbitrary specified Knudsen numbers by expanding the distribution in terms of the Knudsen number\cite{Sone}. However, truncation errors are still inherent to the method. These are also encountered when solving the Boltzmann equation using the Chapman-Enskog expansion, which expands the distribution into an equilibrium and a non-equilibrium part\cite{Chapman}. While this method successfully recovers the Navier-Stokes equations of continuum mechanics, it performs poorly in conditions far from equilibrium. Lastly, the Direct Simulation Monte Carlo (DSMC) method is shown to accurately capture the widest range of Knudsen numbers\cite{Bird1994}. The DSMC method tracks individual molecular motion by advancing particles collisionless, followed by probabilistic collision steps. By simplifying long-range interactions, it is well suited for high Knudsen number flows. Its initial limitation of high computational cost has long been overcome with current implementations fully capable of running on high-performance computing architectures\cite{SPARTA}. Furthermore, its ability to handle complex geometries has established the method as a staple of gas dynamics. 

DSMC has seen extensive developments in the past, and its range of applicability is continually challenged. While methods such as the Chapman-Enskog expansion successfully recover continuum formulations, DSMC shows restrictions for low Knudsen number flows. Resolution of short-range interactions undermine the statistical approach to collisions and higher densities limit the computational advantage it has over other particle methods when computationally parallelized. It is the focus of this paper to improve upon the performance of DSMC by coupling it to molecular dynamics simulations. To achieve this, we implement a method that improves the accuracy of arbitrary regions in the domain by using a cell-based decision technique between the coupling of DSMC and MD. 

\section{Numerical Models}
\subsection{Direct Simulation Monte Carlo}
The Direct Simulation Monte Carlo method excels in large-scale particle simulations. It works by dividing the simulation into three steps. First, all particles undergo Newtonian motion in a freely flowing field. Then, boundaries are taken into account. Reflections, energy transfers, and adsorption may be considered, and surface interactions can be evaluated. Next, collisions are modeled. Rather than using direct atomic interactions, collisions are determined stochastically. Based on parameters such as relative velocity, a fraction of the particles present in a DSMC cell are chosen to collide using an acceptance-rejection scheme. If selected, the particles undergo the interaction based on a pre-determined collision model. Popular choices include the Variable Hard Sphere and the Variable Soft Sphere models\cite{Bird1994}. The Variable Hard Sphere model treats particles as inelastic spheres with an infinite repulsive force upon collision. The Variable Soft Sphere model treats molecules also applies an infinite repulsive force when colliding, but treats the cross-section of particles variable to better match properties such as viscosity and diffusion \cite{VSS1992}. Lastly, macroscopic properties are obtained by evaluating the moments of the distribution functions. 

Due to the stochastic collision operator, DSMC possesses the ability to represent many real particles in the flow by a few simulated particles. The underlying assumption is that particles similar in phase space will undergo similar movement and collisions, such that they can be grouped. The parameter $f_\textrm{num} = \frac{\textrm{physical}}{\textrm{simulated}} \textrm{particles}$ has therefore established itself in the DSMC community. $f_\textrm{num}$ is a major contributor to DSMC's computational efficiency advantage over other particle methods.

The DSMC method was originally designed for rarefied gas simulations by Bird\cite{Bird1994}. It readily captures effects far from equilibrium through its distribution functions. More recently, it has also been applied successfully near continuum conditions such as hydrodynamic instabilities\cite{RMI}\cite{RTI} and turbulence modeling\cite{turbulence1}\cite{turbulence2}. The path to hydrodynamic simulations using DSMC was paved by Garcia, who showed that equilibrium distributions used in continuum mechanics are reached if every flow particle is accounted for in the simulation\cite{Garcia}. Its inherent capability to include compressibility, viscosity, thermal conductivity, and diffusivity may even provide richer insights into hydrodynamic phenomena than continuum methods\cite{RMI}.

The Direct Simulation Monte Carlo method also has significant drawbacks that currently limit its use. In the case of gas-surface interactions, it relies on simplistic models such as the Maxwell reflection model\cite{Maxwell}. Maxwell developed two gas-solid collision cases within kinetic theory: specular reflections conserve particle energy by inverting the velocity's surface normal direction, while diffuse reflections assume that particles reach full thermal equilibrium with the surface and rebound with a Maxwellian velocity distribution at the local surface temperature. Both of these assumptions are often overly simplistic: molecular beam scattering experiments showed a deviation from the Maxwell distribution. An improvement was proposed by Cercignani, Lampis and Lord\cite{CCL}. The CLL model uses accommodation coefficients for the normal and tangential momentum drawn from a well-constructed scattering kernel\cite{gasmodels}. These and many more surface interaction models have gained great popularity because of their simple and efficient implementation as well as accuracy in specific gas flows. Nevertheless, they pose challenges as they often require prior knowledge of the flow species and surface type. They also do not account for changes, and thus do not apply for arbitrary flow conditions. 

Another important limitation of DSMC is its unreliable analysis of low Knudsen number flows. As mentioned, it has been shown to yield good results for hydrodynamic phenomena, especially those affected by non-equilibrium effects in transport properties. For instance, strong shock waves traveling through a fluid can cause instabilities and departure from equilibrium, as observed in the Richtmyer-Meshkov instability (RMI)\cite{Zhou}\cite{Sterbentz}. While it has been shown that DSMC can capture the RMI, its results include inherent statistical noise and are restricted in shock strength\cite{RMI}. In addition, the overly simplistic gas-surface interaction models often fail when dealing with flows at extreme conditions. In such cases, these limitations need to be addressed. The effects driving these deviations from equilibrium often occur on length and time scales much smaller than the captured macroscopic flow field. A multiscale strategy is therefore required.

As outlined above, DSMC displays great accuracy and computational efficiency in many cases. In a few scenarios, it lacks the necessary details for accurate results. These generally occur only within specific time frames and locations. Therefore, DSMC must not be replaced entirely for these situations, but rather complemented by a method which captures the remaining details.

\subsection{Classical Molecular Dynamics}
The DSMC approach relies on analytical and empirical models to describe gas-surface interactions, internal energy modes, ion-kinetic effects and more. These may result in discrepancies for scenarios such as flows in extreme conditions. It therefore proves advantageous to obtain these models from a more accurate representation of particle interactions. The classical molecular dynamics method applies Newtonian equations of motion to model the dynamics of every atom. By tracking the position and velocity deterministically, it creates a highly detailed description of the flow field. Interatomic potentials are used to model the interaction of every atom with its surroundings. Short-range interactions are accurately captured in the classical mechanics framework. In recent works, an increase in range of the interatomic potentials has been achieved, confirming their validity for a variety of conditions\cite{SNAP}. Contrary to DSMC, MD produces continuous atomistic trajectories. Boundary conditions are applied as the atoms advance through space and time. Flow quantities, such as pressure and temperature, are then captured using position and velocity moments.

The molecular dynamics approach provides a clear advantage for gas-surface interactions. The accommodation coefficients (such as those used by the CLL model) can be obtained directly and do not require prior assumptions. The collision between gas atoms and surface atoms can result in adsorption, scattering, or desorption, all of which are readily captured. In addition, the interaction between the gas and the surface influences both the gas and the surface in molecular dynamics. This enables the method to capture realistic surface roughness, surface reactions, and surface temperature changes, which impacts the resulting flow field. Thus, gas-surface interactions could greatly profit from an atomistic description.

Molecular dynamics also provides essential insights to dense flows far from equilibrium. In situations such as strong shock waves, where flow properties drastically change over small scales, molecular dynamics can capture hydrodynamic effects in great detail \cite{RMI-MD}\cite{RMI-MD2}. This influences temperature, viscosity and pressure, among others, which all affect the macroscopic development of the flow field, such as instability growth. In summary, an approach that couples DSMC to molecular dynamics maintains the great advantages of DSMC while leveraging enhanced details of MD.

\subsection{DSMC-MD Coupling}
\begin{figure}
  \centering
  \begin{subfigure}{0.4\textwidth}
    \includegraphics[width=\textwidth]{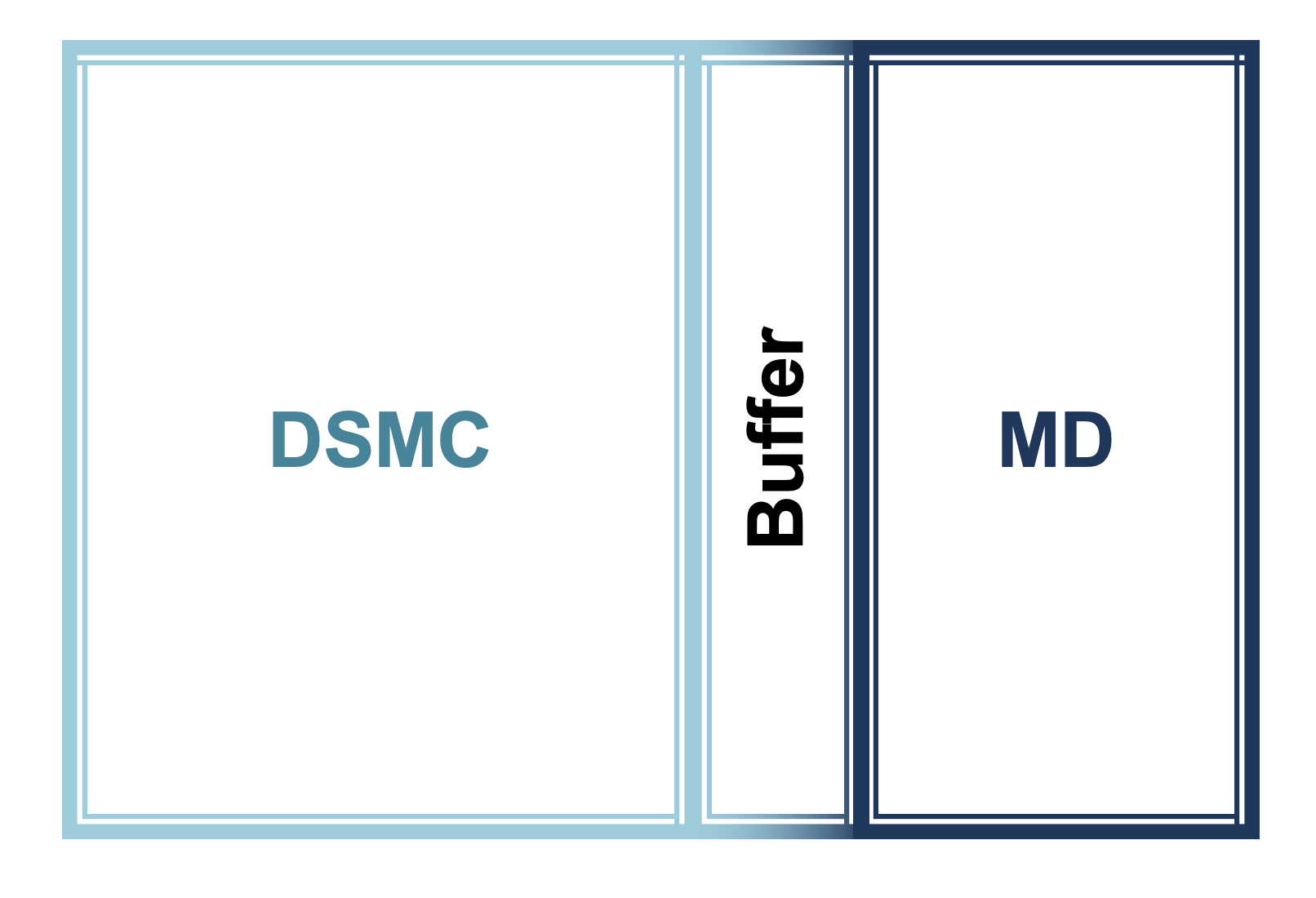}
    \caption{Direct coupling}
    \label{fig:buffer}
  \end{subfigure}
  \hfill
  \begin{subfigure}{0.45\textwidth}
    \includegraphics[width=\textwidth]{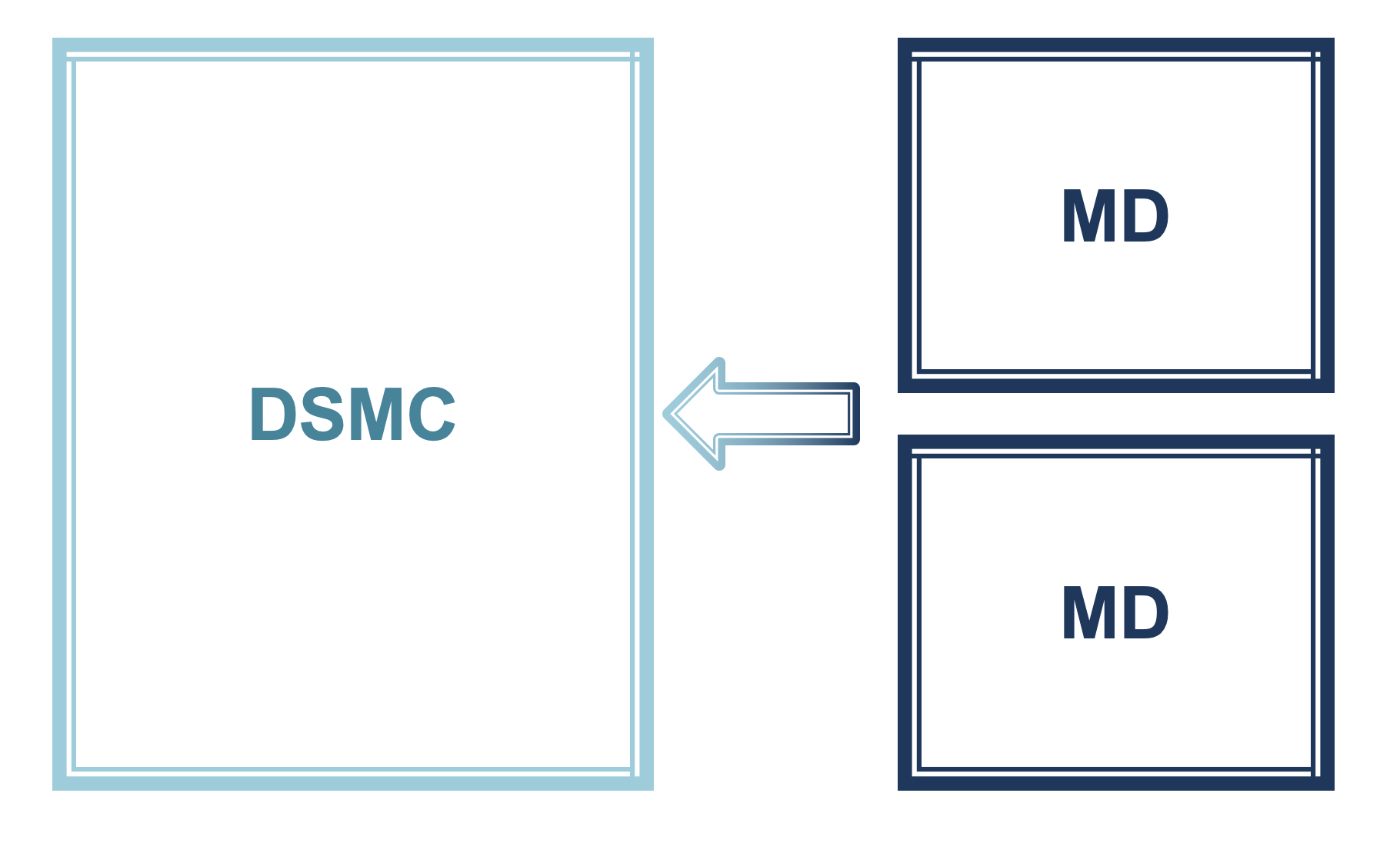}
    \caption{Indirect coupling}
    \label{fig:indirect}
  \end{subfigure}
  \caption{Concurrent coupling approaches. The direct coupling (left) uses an overlapping region between DSMC and MD to transfer information between the two domains. The indirect approach (right) utilizes simulations from MD to support uncertainties in the DSMC domain.}
  \label{fig:couplingApproaches}
\end{figure}

To couple the methods, two main approaches have been established. The buffer zone approach couples the two methods using a jointly overlapping region where both methods meet. It is illustrated in Figure \ref{fig:buffer}. Inside the buffer zone, MD and DSMC particles coexist to transfer information between the two domains. Outside the buffer zone, either the DSMC or the MD method is exclusively used. The buffer zone approach has the advantage of leveraging the computational efficiency of DSMC in the far field, and gaining accuracy at a particular point in space. It also provides a direct physical two-way coupling; i.e., one method directly replaces its counterpart. Particles of each domain are generally treated equivalently, which facilitates information transfer. However, the direct coupling has major disadvantages. Equal treatment of particles in both DSMC and MD neglects DSMC's capability of representing many real particles by a few simulated particles using $f_\textrm{num}$. The buffer zone is also fixed in space, restricting the applicability of the approach as prior knowledge of the coupling location is required. In addition, and perhaps most restrictive, the buffer zone approach requires the DSMC and MD domain to be of similar time and length scales. Nonetheless, multiple implementations have shown successful results using the buffer zone approach\cite{buffer1}\cite{buffer2}\cite{buffer3}\cite{buffer4}.

The indirect coupling method provides an alternative approach, illustrated in Figure \ref{fig:indirect}. Initially motivated to avoid the restriction of matching length and time scales, the indirect approach employs DSMC as the main code and MD for support simulations to supply additional data. Typically, this encompasses the implementation of a database where the results of the molecular dynamics simulations are stored for use by the DSMC method. For example, a variety of velocities and angles of wall incidences may be simulated in the molecular dynamics domain to obtain accurate gas-surface interactions. DSMC particles that experience collisions may then draw from the created database. Another application of the indirect coupling approach is the supplement of specific parameters, such as reflectivity constants, diffusion coefficients, ablation rates, and more. The indirect approach uses both methods to their full capabilities, but restricts itself to flow conditions within the created database. These are often only valid for small systems. Most often, the approach is also one-way coupled: the MD simulation influences the DSMC simulation, but not the reverse. The DSMC simulation has no effect on the MD domain \cite{indirect1}.

The indirect approach is particularly powerful in problems where significant differences in time and length scales between DSMC and MD exist. By passing a determining set of variables from DSMC to MD, support simulations are conducted that quantify the influence of microscopic causes onto macroscopic effects. In addition, the indirect method is naturally scalable due to the physical separation of the two domains. These advantages have been demonstrated in the past \cite{indirect2}\cite{indirect3}.

As shown, the two approaches have found success in their respective range of applications. However, both methods require prior knowledge of the coupling regions or anticipated flow conditions. Therefore, a widely applicable and effective approach to coupling Direct Simulation Monte Carlo and molecular dynamics removes any restrictions on space and flow conditions, and exploits the computational efficiency and physical accuracy of both. This paper presents a novel perspective on achieving this by implementing a concurrent, cell-based coupling method described in the following section.

\section{Cell-Based Coupling Method}
\begin{figure*}
\centering
\includegraphics[width=0.9\textwidth]{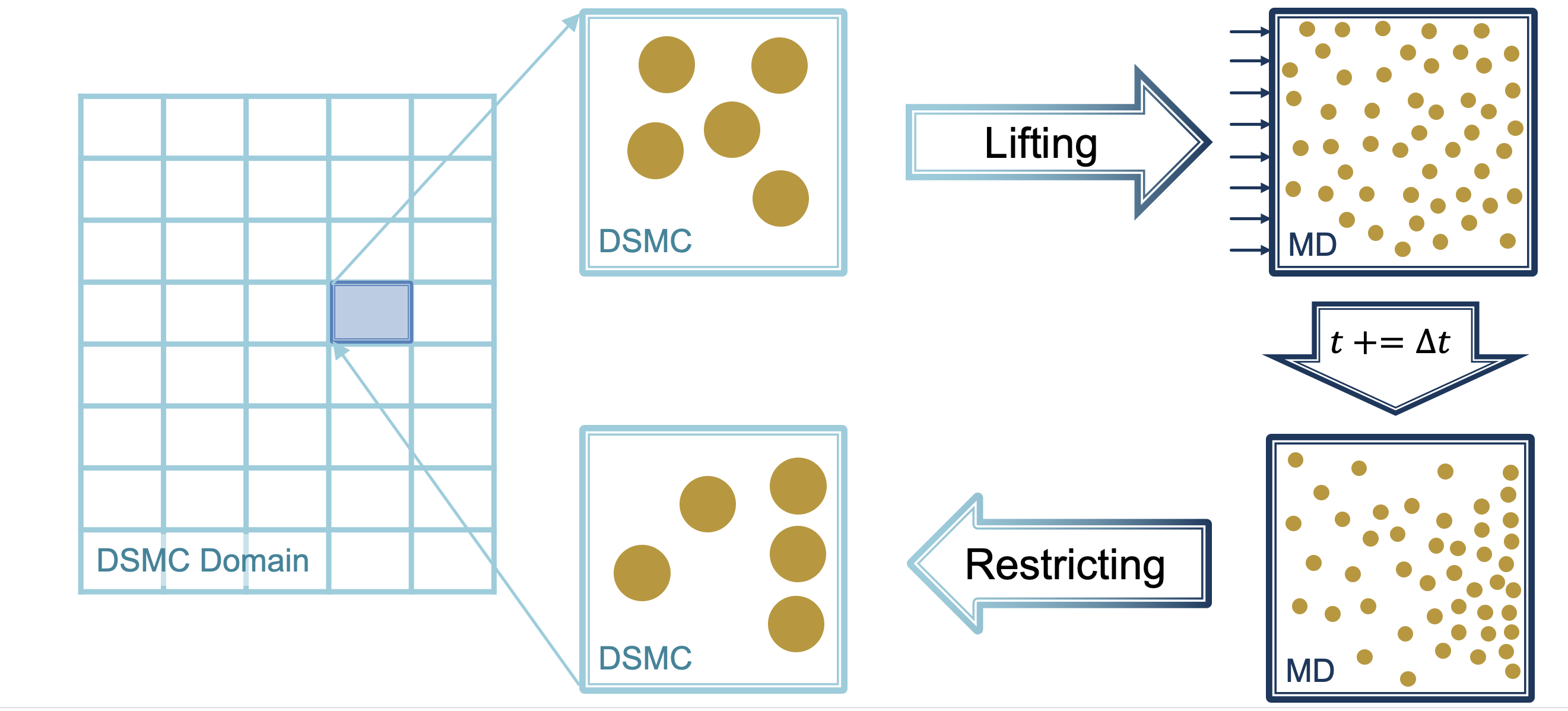}
\caption{Equation-free coupling. The cell-based approach allows MD to supplement constitutive relations to DSMC.}
\label{fig:cellbased}
\end{figure*}

In their essence, multiscale methods attempt to draw macroscopic conclusions from microscopic effects to improve the accuracy of physical models. In the case of DSMC and MD, DSMC is able to span into macroscopic length scales, whereas molecular dynamics excels at capturing the innermost details of a problem. Thus, DSMC is chosen to be the driving method of the simulation, supplemented with data from MD. This idea is in line with the indirect coupling approaches in the literature. A categorization of our approach would fall under the class of heterogeneous multiscale methods\cite{HMM}. Specifically, it follows the general idea of equation-free computation, where microscopic simulations supplement constitutive relations and inform the macroscopic state of a system \cite{equation-free}. Inherently, it lifts the fixed location restriction as well as the length and time scale similarity restriction. To broaden the coupling approach to a wide array of physical processes, a concurrent implementation is chosen. Information desired by the macroscopic simulation (DSMC) is requested and produced immediately by the microscopic method (MD). In this work, we enable the decision to be made on a per-cell basis. When required, one or more cells in DSMC are fully translated into an MD simulation. The cell-based coupling approach is illustrated in Figure \ref{fig:cellbased}.

Two main operators are required to achieve the cell-based coupling approach. The lifting operator transforms the DSMC data into a set of initial conditions for the molecular dynamics simulation. An integration over time in the MD domain follows. The restricting operator then maps the insights from MD back to DSMC. The macroscopic simulation progresses over time and the process repeats. The design of the lifting and restricting operator constitutes the novelty of this paper.

\subsection{The lifting operator}
The concurrent coupling between DSMC and MD must overcome the two major differences between the two methods. First, DSMC does not calculate collisions between particles deterministically. Second, it can represent a group of real atoms by a single simulated particle. The ratio of real atoms to simulated particles $f_\mathrm{num}$ is an essential measure to consider when translating between DSMC and MD. For instance, in the case of a one-to-one ratio, DSMC and MD can be coupled by simply passing the position and velocity of each individual particle between methods. As stated above, this limits the system size within DSMC and requires significant computational resources. Therefore, the lifting operator is tasked with creating a feasible simulation while maintaining a physically accurate system of the DSMC input. 

To fully initialize a simulation in the molecular dynamics domain, we must define all of the following parameters: 1) domain size in x,y,z direction, 2) initial positions of all atoms, 3) initial velocities of all atoms, 4) boundary conditions, 5)interatomic potentials, 6) time integrator.

The translation between domains of these parameters largely depends on the quantities of interest that should be conserved. For particle methods, this is the number density $n$ and velocities, which in turn affect temperature. Velocities are defined per individual particle and atom, and number density is defined as
\begin{equation}\label{eq:nrho}
    n = \frac{N}{V}\ .
\end{equation}
Equation \ref{eq:nrho} shows that if number density $n$ is to be conserved, the volume $V$ of the domain must be adjusted to a changing number of particles $N$. For the following analysis, we will focus on an individual cell of a DSMC domain, translated into an MD system. To reflect a conservation of number density, we must decompose the simulated DSMC particles into their physical particles. Since $f_\mathrm{num}$ is often a very large number ($>10^{12})$, it is obvious that the true $f_\mathrm{num}$ from DSMC cannot be decomposed into  the number of atoms of a feasible molecular dynamics simulation. However, large values of $f_\mathrm{num}$ are a major advantage of the DSMC method, and should not be restricted. Instead, the definition of a new $f_\mathrm{num}$ is required in the MD domain. For this work, the number of MD atoms representing a DSMC particle is called $f_\mathrm{num}^\mathrm{MD}$. We justify this approach by leveraging the periodic boundary conditions in MD, which effectively simulate the dynamics of a much larger system than is explicitly present. This assumption remains valid as long as the number density is conserved across both methods. Different ratios of real to simulated particles also infer that the selected DSMC cell cannot be translated directly into the molecular dynamics domain. This proves advantageous as we do not pose any restrictions on the size of a DSMC cell with respect to the feasibility of a molecular dynamics simulation. 

We scale the DSMC cell to conserve its number density in line with Equation \ref{eq:nrho}. For the remainder of the work, all quantities lacking a superscript refer to the DSMC domain, and all MD attributes are labeled with a superscript. Given the DSMC cell coordinates $[x_\mathrm{lo}, y_\mathrm{lo}]$ and $[x_\mathrm{hi}, y_\mathrm{hi}]$ denoting the lower and upper bounds in two dimensions, we obtain the aspect ratio of the cell:
\begin{equation}\label{eq:AR}
    AR = \frac{x_\mathrm{hi} - x_\mathrm{lo}}{y_\mathrm{hi} - y_\mathrm{lo}}\ .
\end{equation}
Using Equation \ref{eq:nrho}, we find the required relation of the MD domain to conserve number density for a given number of DSMC particles $N$ and a chosen constant $f_\mathrm{num}^\mathrm{MD}$.
\begin{align}
    n &= \frac{f_\mathrm{num}N}{V} = \frac{f_\mathrm{num}^\mathrm{MD}N}{V^\mathrm{MD}}\\
     &= \frac{f_\mathrm{num}^\mathrm{MD} N}{(x\cdot y\cdot z)^\mathrm{MD}}\ .
\end{align}
Conserving both the original aspect ratio of the DSMC cell as well as the number density, we obtain the new MD domain size for the MD domain, from which we derive the individual lengths $x^\mathrm{MD}$ and $y^\mathrm{MD}$
\begin{align}
    (x y)^\mathrm{MD} &= \frac{f_\mathrm{num}^\mathrm{MD} N}{n \cdot z^\mathrm{MD}}\\
    x^\mathrm{MD} &\stackrel{(\ref{eq:AR})}{=} AR \cdot y^\mathrm{MD}\label{eq:xMD} \\
    y^\mathrm{MD} &= \sqrt{\frac{f_\mathrm{num}^\mathrm{MD} N}{AR\cdot n\cdot z^\mathrm{MD}}} \ . \label{eq:yMD}
\end{align}

Note that we assume the MD domain to be initialized through the origin, and that the representation is restricted to rectangular meshes, a common choice in DSMC. The geometry of the molecular dynamics domain is initialized with the updated size. This proves a match of number density across the two domains.
\begin{equation}
\begin{split}
    n^\mathrm{MD} &= \frac{f_\mathrm{num}^\mathrm{MD} N}{(x\cdot y\cdot z)^\mathrm{MD}}\\
    &\stackrel{(\ref{eq:xMD})}{=} \frac{f_\mathrm{num}^\mathrm{MD} N}{AR (y\cdot y\cdot z)^\mathrm{MD}}\\
    &\stackrel{(\ref{eq:yMD})}{=} \frac{f_\mathrm{num}^\mathrm{MD} N}{AR\cdot z^\mathrm{MD} \frac{f_\mathrm{num}^\mathrm{MD} N}{AR\cdot n\cdot z^\mathrm{MD}}}\\
    &= n \ .
\end{split}
\end{equation}

Next, particles must be initialized. While DSMC tracks the position and velocity of each of its particles, which can be easily translated to MD, it becomes nontrivial for cases of $f_\mathrm{num} > 1$. Here, many particles are represented by one position and velocity, which is highly unphysical if translated to the molecular dynamics domain. Two particles that share the exact same position would be subject to an infinite repulsive force in most interatomic potentials, so care must be taken to avoid energy artifacts when placing particles. This could be done through an energy minimization step, which iteratively places particles until an energy minimum is found. However, minimizing thousands of exactly overlapping particles is highly inefficient and often infeasible. Alternatively, particles can be randomly distributed across the simulation domain. This mostly avoids spatial overlap, but undermines the concept of the $f_\mathrm{num}$ ratio: DSMC can represent many physical particles by one simulated particle if they are assumed to be very similar in phase space. Should their positions be dispersed, this no longer holds true and important aspects such as non-equilibrium effects may be lost. To ensure that phase space is conserved in the molecular dynamics simulation, and energy artifacts are avoided, a more complex initialization of particle positions is required. 

\begin{figure}
\centering
\includegraphics[width=0.4\textwidth]{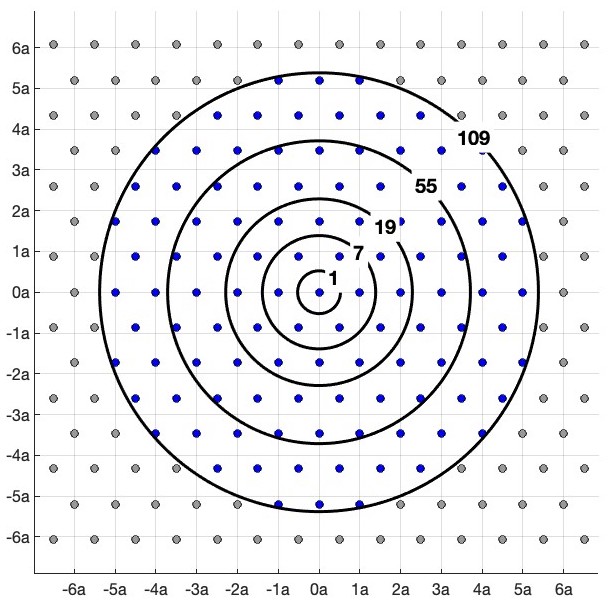}
\caption{Decomposition of DSMC particles to MD atoms. Atoms are initialized in a circular region across a close-packed hexagonal lattice.}
\label{fig:latticecircles}
\end{figure}

With a set domain size of a given $f_\mathrm{num}^\mathrm{MD}$, $N$, and $AR$, the lifting operator must ensure that the decomposition of simulated DSMC particles into $f_\mathrm{num}^\mathrm{MD} N$ atoms positions them closely. When minimizing the distance among a group of atoms, one naturally obtains a hexagonal close-packed (hcp) lattice structure. All major molecular dynamics codes support the initialization of atoms into an hcp lattice, and the lifting operator takes advantage of this. To maintain the DSMC's particle original form of a circle, but allow for a variable choice of $f_\mathrm{num}^\mathrm{MD}$, the initialization occurs in a circular region of variable radius. To optimally choose the radius of the region, we consider the problem of determining how many lattice points lie in a circle of radius $r$. This problem has been extensively studied for square lattices, known as the Gauss Circle problem, and was later solved for Euclidean as well as Non-Euclidean spaces: Lax \& Phillips \cite{Lax1982} found that for an arbitrary space, the number of lattice points $N(r)$ depends on the radius approximately as 
\begin{equation} \label{eq:hexlattice}
    N(r) \approx \frac{\pi r^2}{|F|},
\end{equation}
where F is the volume of the fundamental domain of the crystallographic group. Given an hcp 2D lattice with lattice constant $a$, its unit area is mapped by the coordinates $(a,0)$ and $(\frac{a}{2}, \frac{\sqrt{3}a}{2})$. Thus, the unit area containing three points is $|F| = \frac{\sqrt{3}}{2}a^2$. Equation \ref{eq:hexlattice} becomes
\begin{equation}
    N(r) \approx \frac{2\pi r^2}{\sqrt{3}a^2}\ .
\end{equation}
This consequently leads to an expression for the radius which encloses $N$ atoms in a lattice of spacing $a$:
\begin{equation} \label{eq:latticeRadius}
    r \approx \sqrt{\frac{\sqrt{3}N}{2\pi}}a \ .
\end{equation}
For a given number of atoms that represent each DSMC particle, Equation \ref{eq:latticeRadius} provides the required radius by which they are translated into the molecular dynamics domain. Although this progress is only approximate (with a finite error that is known), it avoids the necessity to place individual atoms in the MD domain, a task that would otherwise be computationally prohibitive. Figure \ref{fig:latticecircles} shows the circles resulting from Equation \ref{eq:latticeRadius}, labeled by their number of enclosing points. The formulation holds for all values of the lattice spacing $a$.

\begin{figure}
\centering
\includegraphics[width=0.6\textwidth]{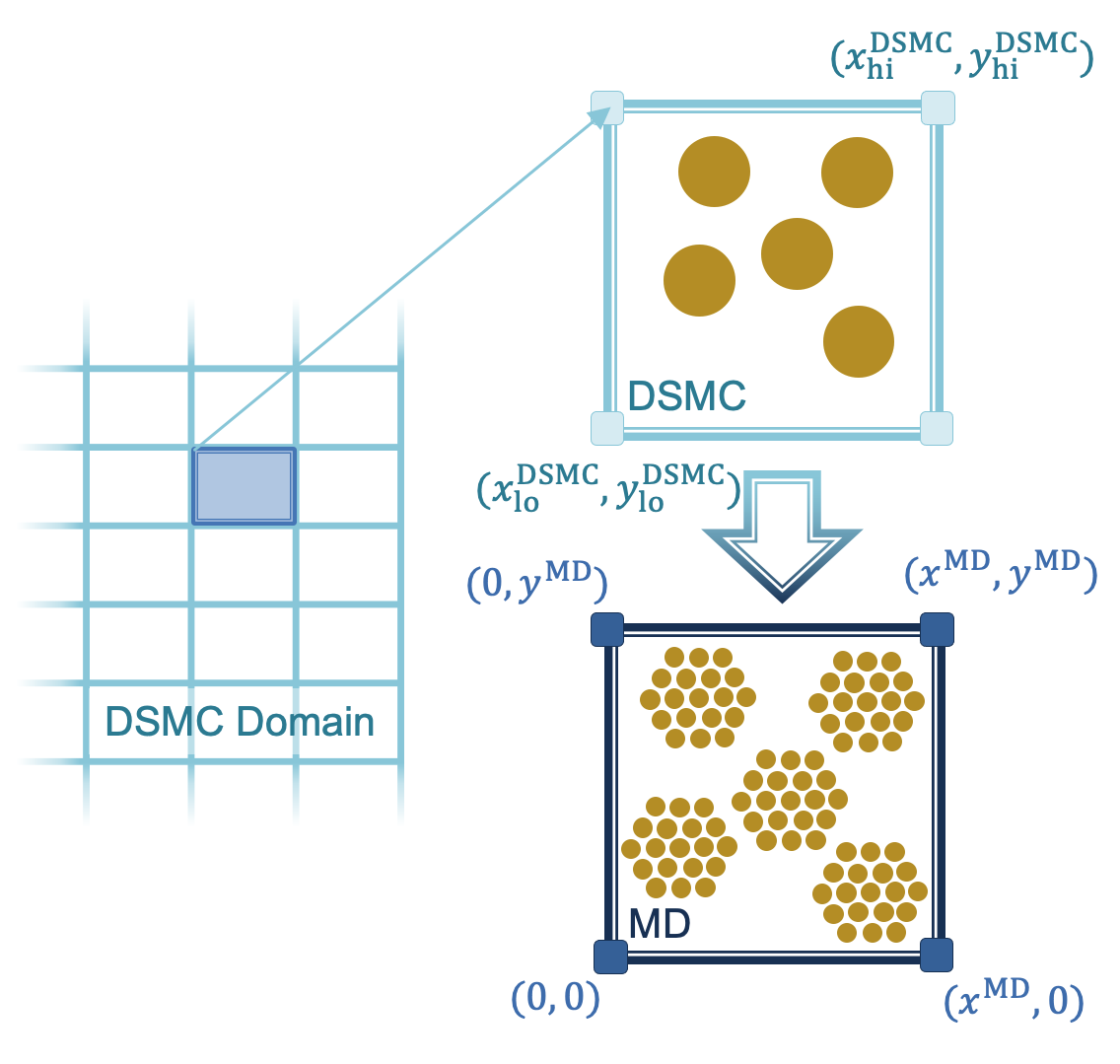}
\caption{The lifting operator. The DSMC geometry and its particles are mapped to an initial MD domain with atoms.}
\label{fig:lifting2}
\end{figure}

The lattice spacing has a direct influence on the configurational energy of the system, and should be chosen accordingly. To minimize the energy but maintain a minimal distance between atoms, the interatomic potential is considered. It takes in the distance between two atoms and returns a force acting on each one. True potential formulations possess an infinite range, which makes the MD computation very inefficient when considering all interactions between atoms at any distance within the considered domain. In practice, interactions are restricted to certain lengths. The corresponding distance is the cut-off distance $r_c$. Two atoms more than $r_c$ apart are unaffected by each other's presence, and do not add energy to the domain. Thus, an energy minimization can be achieved by placing atoms within a cut-off distance from each other. Choosing the lattice spacing of the hexagonal close packed lattice to be $a = r_c$ therefore ensures an energy minimum. For the marginal case that two DSMC particles are close enough such that their resulting lattice structures overlap in the MD domain, one of the overlapping atoms is deleted. The domain and position initialization is exemplified in Figure \ref{fig:lifting2}. For the purposes of the illustration, it was chosen that $f_\mathrm{num}^\mathrm{MD} = 19$.

To complete the list of required MD definitions, the lifting operator translates information regarding boundary conditions, interatomic potentials and time integrators. These three items are unique to the use case at hand, and cannot be universally implemented. Sections \ref{sec:verification} and \ref{sec:example} provide two examples of defining these items.

To summarize, the lifting operator is called at the end of each DSMC time step. Based on a user-given input or criterion, the operator automatically recognizes which cells are chosen to be coupled to molecular dynamics. For instance, gas-surface interactions, phase boundaries or regions of highly non-equilibrium behavior may be selected. The chosen cells can be all or any subset of the DSMC simulation cells. The domain size is translated from DSMC to MD, and positions and velocities are set. Boundary conditions are selected, interatomic potentials are defined and the time integrator(s) are chosen. The molecular dynamics simulation can perform a detailed analysis, whose result is provided to the next DSMC time step. The restricting operator is then responsible for mapping back the findings.

\subsection{The restricting operator}
After running the molecular dynamics simulation, the new results shall be used to inform the Direct Simulation Monte Carlo domain. All particle simulations are defined through their mass, position, and velocity, including MD and DSMC. The restricting operator must carefully choose MD atoms and process their information to include in the DSMC domain. To maintain the same mass of one DSMC particle, the same number of MD atoms is grouped to form a DSMC particle as those initially created during the lifting step. Note that the term mass is used loosely here and does not imply an addition of MD atoms to one big particle. Rather, it denotes the center-of-mass and reflects that the influence of each MD atom should be equally weighed in the DSMC simulation. A mismatch in the number of MD atoms decomposed from DSMC in the lifting operator and later regrouped in the restricting operator would unphysically skew the results. 

\begin{figure}
\centering
\includegraphics[width=0.6\textwidth]{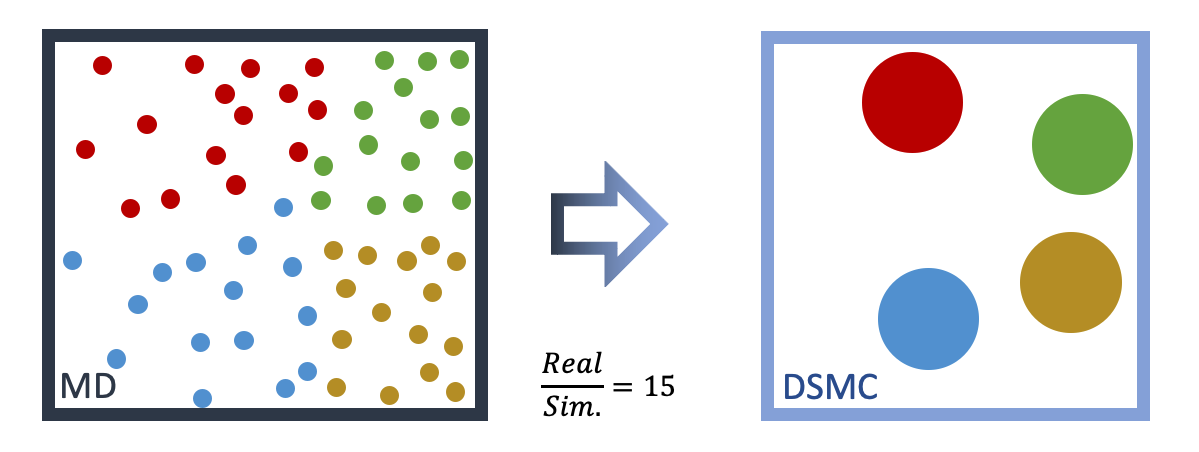}
\caption{The restricting operator. Atomistic insights gained in the MD domain are provided to the DSMC simulation.}
\label{fig:restricting}
\end{figure}

From the molecular dynamics simulation, the position and velocity of each atom is listed. A sorting algorithm is applied that orders the list according to the nearest neighbor criterion. The authors recognize the computational expense of the nearest neighbor sort ($\mathcal{O}(n^2)$), and tested a variety of less expensive sorting criteria, e.g. w.r.t. separate x and y positions. We came to the conclusion that the physical inaccuracies involved with an unfit sorting outweigh the computational gain, especially in light of the overall cost of a full molecular dynamics simulation needed to provide the list. However, improvements to this could be made by using grid-based hashing approaches or particle cell structures\cite{Beazley1994}. To form each nearest-neighbor group, exactly $f_\mathrm{num}^\mathrm{MD}$ atoms are considered. The grouped atom positions and velocities are averaged and assigned to the updated DSMC particles. Figure \ref{fig:restricting} shows an example of how the restricting operator would behave in the case of $f_\mathrm{num}^\mathrm{MD} = 15$. 

Inherent to any translation from a more detailed to a simpler description, not all information can be used. In the case of the restricting operator, the distribution of particle positions and velocities is simplified to a single position and average velocity. This is designed such that a DSMC particle still represents a slice of phase space used in the Boltzmann formulation. Thus, using an average rather than the distribution of the many MD particles causes a loss of information. In contrast, the lifting operator introduced additional information to the simulation, sourced from assumptions not drawn from DSMC. The restricting step counteracts the addition of that information. While the initial position distribution of the MD domain is initialized as a lattice, the restricting operator reduces the distribution to a single particle. And while the initial velocity distribution closely represents phase space with the addition of thermal fluctuations, the restricting operator ensures that phase space is truly conserved. Any artifacts which may have resulted from the molecular dynamics simulation are most appropriately accounted for. 

\section{Verification of the Method}\label{sec:verification}
The lifting and restricting operators are implemented using open-source software libraries. For the Direct Simulation Monte Carlo method, the Stochastic PArallel Rarefied-gas Time-accurate Analyzer (SPARTA) is selected \cite{SPARTA}. For the molecular dynamics software, the Large-scale Atomic/ Molecular Massively Parallel Simulator (LAMMPS) is used \cite{LAMMPS}. The functionality of the method is verified through confirmation of the applicable conservation laws. For this, we design a very simple simulation of atomic oxygen with periodic boundaries. Three cells are randomly selected to be fully coupled to MD at every DSMC time step. The molecular dynamics simulation automatically receives its domain size, atom positions and velocities from SPARTA. It runs using periodic boundary conditions and a simple Lennard-Jones potential for oxygen. It performs its simulation in the microcanonical (NVE) ensemble.

\subsection{Conservation of Mass}
\begin{figure*}
\centering
\includegraphics[width=0.99\textwidth]{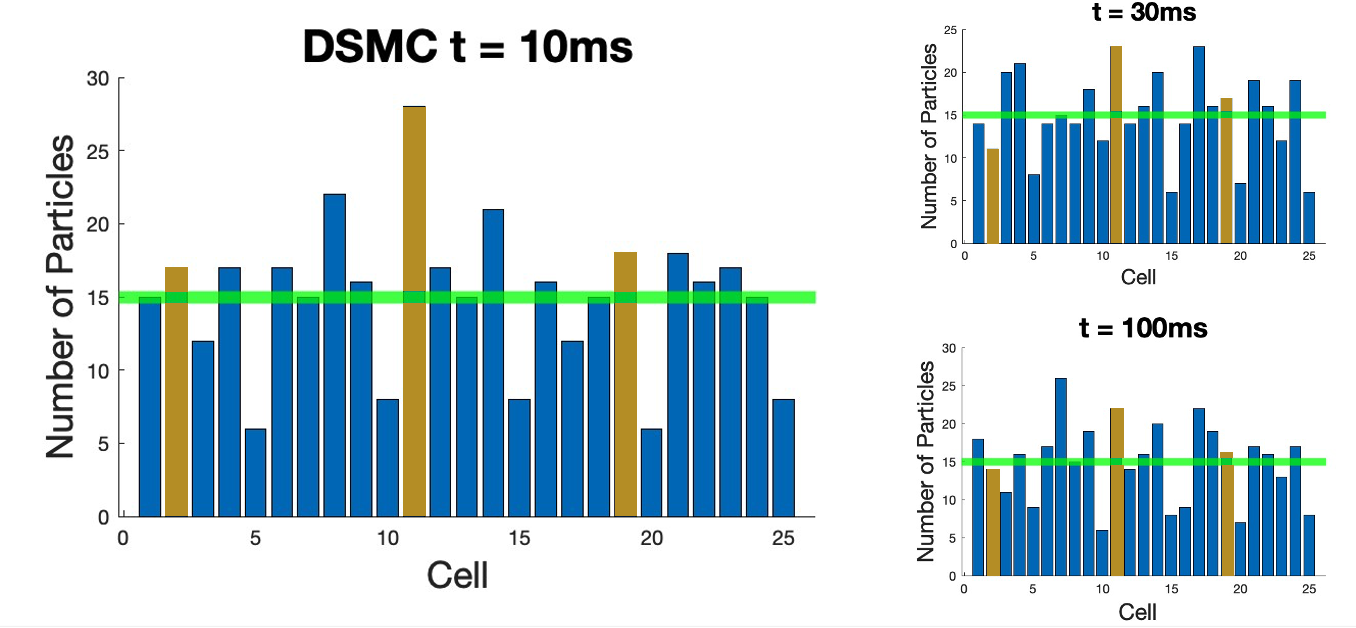} 
\caption{DSMC particle count per cell. Fluctuations in number density are observed for DSMC (blue) and DSMC-MD cells (gold) with an obtained average across all cells (green) for multiple time steps.}
\label{fig:mass}
\end{figure*}

To investigate the validity of the information exchange between DSMC and MD, we first observe how the coupled cells interact in comparison to non-coupled DSMC cells. Over the span of 100 milliseconds, we observe a natural progression of particle movement across cells. On average, 15 particles per cell are expected. The system was chosen particularly small both in number of cells and particle count to expose any irregularities in the distributions. Figure \ref{fig:mass} shows the particle distribution per cell of the simulation. The golden distribution bars denote the coupled cells. Samples are taken at $t=10, 30$ and $100$ milliseconds to provide two consecutive distributions shortly after initialization ($t=0$ would give a trivial, close to uniform distribution of the average) and a distribution of the end of the run. During each time step, the lifting and restricting operator perform a coupled simulation. No particular bias or anomaly is observed, and changes in the distribution correspond to the expected levels of such small systems. 

We observe a consistent average of 15 particles per cell across the three plotted time steps. Figure \ref{fig:mass} confirms that the average number of particles per cell in the domain remains constant across all time steps. It is clear that there are no changes in mass in the domain, which indicates that number of atoms initialized by the lifting operator are correctly grouped by the restricting operator. These results firmly prove a conservation of mass of the cell-based coupling approach.
\begin{figure*}
\centering
\includegraphics[width=0.99\textwidth]{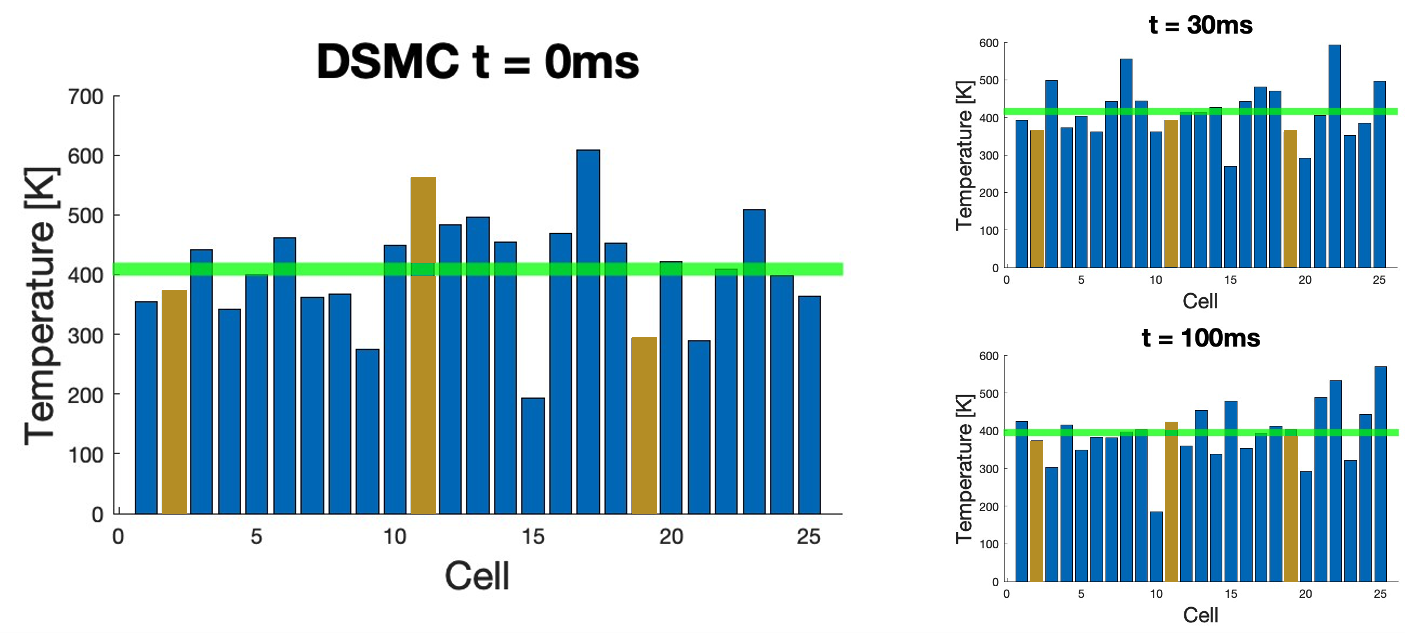} 
\caption{Temperature per cell. Fluctuations are observed in DSMC (blue) and DSMC-MD cells (gold) with an obtained average across all cells (green) for multiple time steps.}
\label{fig:temp}
\end{figure*}
\begin{figure*}
\centering
\includegraphics[width=0.99\textwidth]{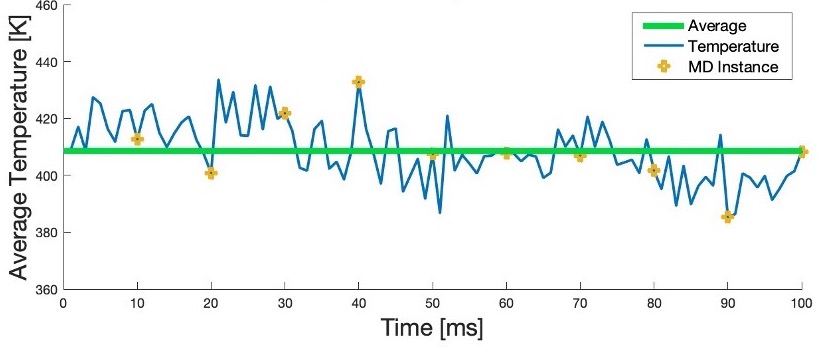} 
\caption{Average temperature in DSMC. The measured temperature is observed over time for a DSMC simulation coupled to MD.}
\label{fig:conservation-temp}
\end{figure*}

\subsection{Behavior of Temperature}
The lifting and restricting operators are tested with respect to their treatment of velocities. Being a particle-based simulation, DSMC derives all higher moments from its velocity distribution, such as temperature. Therefore, the behavior of the temperature is observed. 

In Figure \ref{fig:temp}, we show the distribution of average temperature in the DSMC domain. The golden bars denote the coupled cells, and the green bar indicates the average temperature. Samples are taken at $t=0, 30$ and $100$ milliseconds to provide the initial distribution, one shortly after initialization and the distribution of the end of the run. During each time step, the lifting and restricting operators perform a coupled simulation. Once more, the coupled cells show no favoritism compared to other cells to drain or add energy into the system, underlining the success of the lifting and restricting operators. The fluctuations across different cells correspond to the expected uncertainties of such a small system.

While the temperature averages across the entire simulation domain show qualitative agreement, a more detailed evolution is shown in Figure \ref{fig:conservation-temp}. To unveil any inaccuracies with the coupling method, we ran the simulations with a coupling of DSMC-MD for every cell every 10 timesteps. Across all time steps, a direct comparison between fluctuations caused by the MD instance on non-coupled time steps is drawn. No particular bias is observed by the MD instance. Given the small deviations from the mean, we conclude that these fluctuations are due to the inherent properties of finite size systems found in the DSMC runs. In fact, we observe that these finite size effects conform to the statistical mechanics of small systems. The standard deviation of the coupled DSMC-MD fluctuations is $\sigma = 10.56$ Kelvin. The theoretical standard deviation is calculated using
\begin{equation}\label{eq:stddvn}
    \sigma = \sqrt{\frac{2T_0^2}{3N_0}},
\end{equation}
where $T_0$ is the average temperature and $N_0$ the average number of particles\cite{Hadji2003}. Note that we assume the gas to be a monatomic ideal gas, and therefore employ the heat constant of a particle at constant volume as $c_V = \frac{3}{2}k_B$. For an average temperature of 409.1 Kelvin and 375 particles, Equation \ref{eq:stddvn} provides that the standard deviation should be on the order of $\sigma = 17.25$ Kelvin, which is reasonably close to our findings. The fact that the standard deviation in the DSMC-MD coupled simulation slightly outperforms the predicted theoretical value can be attributed to the replacement of the stochastic collision operator in DSMC by deterministic collisions in MD, eliminating a source of additional statistical variance. The finite size effects are expected to be larger in DSMC compared to MD, since the coarser model has less degrees of freedom. The exact influence of an improved statistical behavior in the DSMC-MD coupling is subject to further investigation. The general match in deviations from the average argues that the lifting and restricting operators possess no inconsistencies and are replicating the correct temperature behavior within their range of statistical plausibility.


\section{Application} \label{sec:example}
At hypersonic conditions, the flow around an object or vehicle causes extreme temperature and density gradients. Specifically at the surface, it is essential to accurately capture temperature to predict heat transfer to the payload, reactivity of the surface and subsequent ablation of the body. Accurately predicting the surface temperature evolution is therefore of primary importance when simulating an object in a hypersonic flow. In experiments, the temperature of a hypersonic flow through a nozzle is often measured using a thermocouple. The thermocouple is introduced into the flow as a thin wire that is far smaller than the flow field dimensions. We illustrate the usability of the general cell-based coupling method using a complex gas-surface interaction between a thermocouple and a hypersonic flow environment. 

\begin{figure}
\centering
\includegraphics[width=0.6\textwidth]{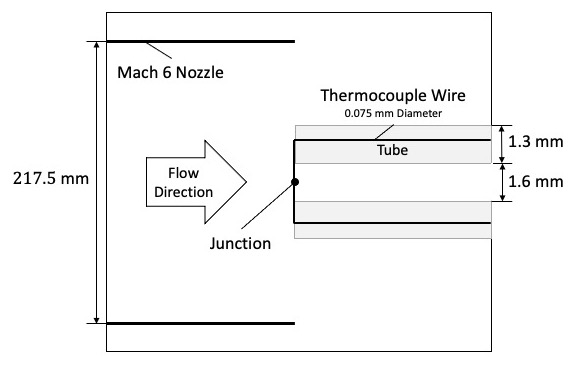} 
\caption{Experimental Set Up of Stagnation Temperature Measurements in a Hypersonic Flow (Not to Scale) \cite{Wire2013}.}
\label{fig:experimental-setup}
\end{figure}

In a series of experiments, Widodo \& Buttsworth \cite{Wire2013} obtained temperature readings for a cold hypersonic flow. Their experimental set-up is shown in Figure \ref{fig:experimental-setup}. Conducted at the University of Southern Queensland hypersonic wind tunnel facility\cite{Buttsworth2009}, which is well-suited for cold flows at relatively long test times, the tests featured free piston compression heating. Temperature measurements were taken using a 0.075 mm wide k-type butt-welded thermocouple wire positioned in the test section of a Mach 6 hypersonic nozzle and exposed to the flow for around 200 ms. The nozzle through which the flow was accelerated had a diameter of 0.2175 m. Such experiments are particularly useful for predicting the overall heat load a hypersonic body may be exposed to. They also assist the validation of numerical simulations. To model the set up, a simulation must be able to capture both the flow in the nozzle as well as the impact of particles hitting the small wire. The large difference in length scales between the larger flow field and the thermocouple ($0.2175$ m compared to $7.5 \cdot 10^{-5}$ m) may demand a multiscale approach. Our concurrent DSMC-MD coupling framework is therefore well positioned to capture the relevant phenomena, and we present the performance of our method compared to a conventional DSMC simulation.

\subsection{Simulation Set Up}

The DSMC simulation was set up at the specified conditions of an experimental run. Widodo \& Buttsworth \cite{Wire2013} measured stagnation pressure $P_0$, stagnation temperature $T_0$, ambient temperature and the initial wire temperature for a Mach number of $Ma = 5.84$. The conditions are summarized in Table \ref{table:conditions}. These properties are translated to flow properties relevant to DSMC. Through the isentropic relations, we obtain the flow conditions at the exit of the nozzle $T_\textrm{exit} = 73.14$ K and $P_\textrm{exit} = 687.62$ Pa. Assuming an ideal gas, the density equals $\rho=0.033 \frac{\textrm{kg}}{\textrm{m}^3}$ which translates to a number density of $n = 6.809 \cdot 10^{23} \frac{1}{\textrm{m}^3}$. The gas flow velocity results in $V = Ma \cdot \sqrt{\gamma RT} = 1001.14 \frac{m}{s}$. For the DSMC simulation, these conditions are assumed to hold constant across time and space at the inlet boundary.

\begin{table}[t]
\centering
\begin{tabular}{l l}
  Parameter & Value \\
  \hline
  Ma & 5.84 \\ 
  $P_0$ & 0.920 MPa \\
  $T_0$ & 572.0 K \\
  $T_\textrm{amb}$ & 298.0 K \\ 
  $T_\textrm{init,wire}$ & 541.0 K \\
\end{tabular}
\caption{Experimental Conditions \cite{Wire2013} After Which the Simulations Are Modelled.}\label{table:conditions}
\end{table}

The far field of the simulation is chosen to extend $4.0$ mm above and below the tube housing the thermocouple, such that macroscopic field effects such as shocks are captured. $10$ mm in front and $2.2$mm aft of the wire junction are considered. This domain is divided into grid cells. The choice of number of grid cells is intrinsically coupled to the number density, the ratio between real and simulated particles ($f_\mathrm{num}$), geometric considerations and other parameters. Ultimately, the number of grid cells is chosen such that sound statistical quantities can be obtained, which is typically the case for at least 10 simulated particles per grid cell \cite{Shu2004}. For the simulational set up of the hypersonic flow, this results in a 21 by 21 mesh with $\Delta y = \Delta x = 0.6$ mm, a typical cell size for DSMC simulations. Since $\Delta y = \Delta x > d_\textrm{wire}$, the wire is captured entirely within a single grid cell. The simulation is run in 2D with an $f_\mathrm{num}=5\cdot10^{16}$ and a time step of $\Delta t = 10^{-5}$ s. The DSMC time step is chosen sufficiently small such that particles do not traverse an entire cell within one time step. The boundary conditions are selected to be reflective in y-direction and open in x-direction. The tube housing the wire is placed at the end of the rightmost end of the domain. The gas is comprised of $\textrm{O}_2$, $\textrm{N}_2$, O, N and NO. A variable soft sphere model is used for collisions. For a detailed view of the simulation domain, we refer the reader to a video linked in the supplemental materials.

For a standard DSMC simulation, we implemented a surface temperature model at the surface of the thermocouple wire junction. Using the flux $q_\mathrm{surf}$ of total energy onto the surface element, the temperature of each surface element is calculated from the Stefan-Boltzmann law for a gray-body\cite{SPARTA}:
\begin{equation}\label{eq:stfboltz}
    T_\mathrm{surf} = \sqrt[4]{\frac{q_\mathrm{surf}}{\sigma \epsilon}},
\end{equation}
where $\sigma$ denotes the Stefan-Boltzmann constant and $\epsilon$ the emissivity. Provided the conductivity of the wire \cite{Wire2013}, the emissivity can be calculated according to Atallah's relationship for metals\cite{Atallah1966}:
\begin{equation}\label{eq:emissivity}
    \epsilon = 10^{-4} T k^{-\frac{1}{2}},
\end{equation}
where $10^{-4}$ is a semi-theoretical prefactor, $k$ denotes the conductivity and $T$ the temperature. Widodo \& Buttsworth \cite{Wire2013} provide the wire's conductivity as $k = 0.25 \frac{\textrm{W}}{\textrm{cm K}}$, such that $\epsilon = 10^{-4}\cdot 541 \cdot (0.25)^{-\frac{1}{2}} = 0.1082$. Thus, particles impacting the thermocouple cause a temperature change of the wire, which in turn updates the diffuse surface collision model.

We postulate that DSMC will encounter limitations with its surface model. Due to the large differences in spatial scales, Equation \ref{eq:stfboltz} will likely under/over-estimate the effect of the flow on the wire. Therefore, we replace the surface model with the concurrent multiscale framework. The grid cell containing the wire element is selected to be fully coupled to molecular dynamics.

Within the MD domain, atom positions and velocities are automatically defined by the lifting and restricting operators. The domain size is translated according to Equations \ref{eq:xMD} and \ref{eq:yMD}. Boundary conditions are set as periodic in y-direction. The inflow boundary in the x-direction is set as a specular reflection, indicating that for every flow atom that would exit, an exact replica would enter the flow domain, ensuring conservation of number density. The outflow in the x-direction is filled with layers of nickel atoms, the constituents of the wire. Its boundary condition is a wall whose atomic interactions match the interatomic potential of nickel, effectively extending the wire width indefinitely. Thus, the procedure successfully bridges the length scales from macroscopic to atomistic. The flow is made of atoms whose attributes are averages over all air constituents. For nickel-nickel interactions, the embedded atom potential designed by Stoller et al.\cite{NiEAM2016} is used. The average interaction parameters for air in form of a Lennard-Jones potential are provided by Chen \& Yuen \cite{LJair2010}. The interaction between nickel and air atoms are modeled as a Lennard-Jones potential with parameters according to the Lorentz-Berthelot combining rules \cite{Lorentz}\cite{Berthelot}.

\begin{figure}
\centering
\includegraphics[width=0.4\textwidth]{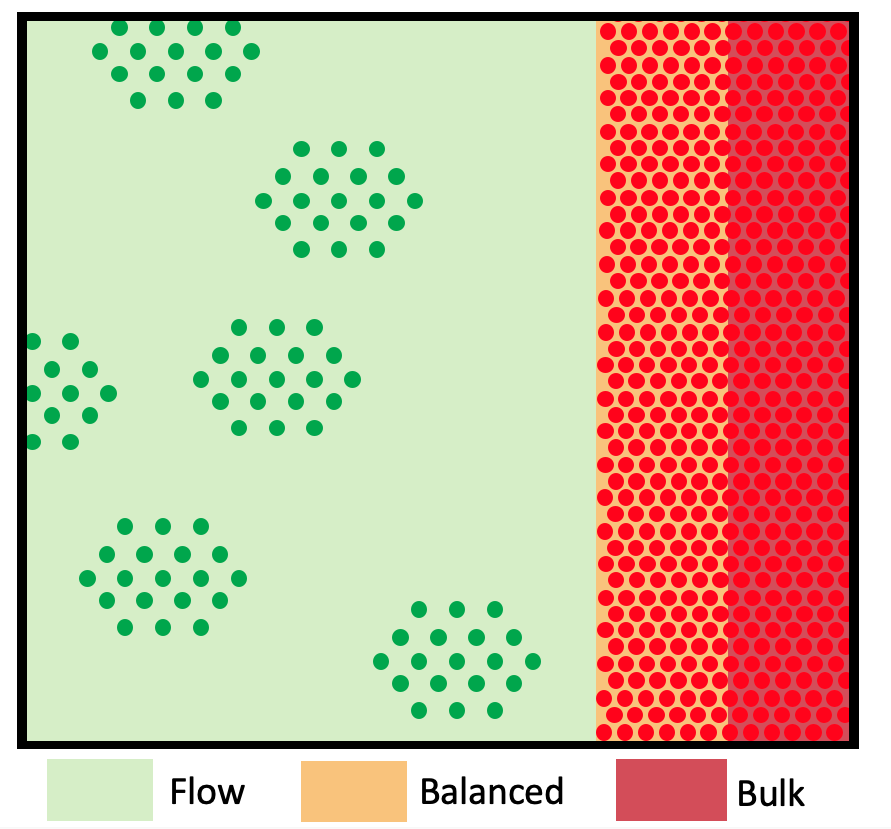} 
\caption{The molecular dynamics domain. Air (green) and nickel (red) atoms are divided into three regions to capture transient effects\cite{Wire2013}}.
\label{fig:md-domain}
\end{figure}

To capture the heating and cooling of the wire in MD accurately, we balance the effect of impeding atoms onto the surface with the radiative heat loss experienced by the wire. The MD domain is divided into three regions: the flow region, a balanced layer and a bulk layer. Figure \ref{fig:md-domain} shows the flow region, the balanced region as well as the bulk region. In the flow region, the air atoms are initialized by the lifting operator. The time integration is run in the microcanonical ensemble. The bulk region houses wire atoms which are assumed to possess steady-state properties throughout the simulation. The balanced region is a layer spanning multiple nickel lattice widths between the flow region and the bulk region. It is in direct contact with the flow as well as the bulk wire atoms, such that the influence of both is reflected in its atoms. It is within this region that the surface temperature is measured. Its thickness is chosen such that sufficient interatomic collisions occur to capture the gradient between the flow and bulk regions. According to the Knudsen number concept of non-continuum behavior, this is the case for a thickness of at least 10 atoms if the gradient shall be captured across the width. To accurately capture momentum transfers with the flow, the atoms in the balanced region are also run in the microcanonical ensemble. The bulk region in the wire is thermostatted in the canonical ensemble using a Nose-Hoover thermostat \cite{LAMMPSnvt}. The balance between atom-surface interactions and radiation heat loss is designed to emulate the DSMC surface model closely to enable a direct comparison.

The molecular dynamics simulation is then run. First, all wire atoms (in both the temperature measurement region and the heat source region) are thermostatted until the desired temperature is reached. A small time step of $\Delta t = 5\cdot 10^{-16}$ s is used to avoid temperature artifacts. We equilibrate the system for a duration of $10^4$ timesteps. During this time, flow atoms are not integrated. Once the wire atoms have the desired target temperature, collisions between flow atoms and wire atoms are captured. Due to the high-energy collisions between air and wire atoms, the MD simulation is conducted using $\Delta t = 10^{-15}$ s. The MD time step is chosen such that high speed interactions between nickel and air atoms are accurately captured. The simulation is run until an equilibrium in temperature evolution is reached, and the temperature is averaged over the last 20,000 timesteps. The resulting temperature is passed back to the DSMC domain. DSMC particles are updated by the restricting operator. The DSMC simulation continues with its updated surface temperature and particles.

With spatial scales bridged, the problem of temporal discrepancies must be overcome. The DSMC domain runs with a timestep of $\Delta t = 10^{-5}$ s, while the MD domain uses a timestep of $\Delta t = 10^{-15}$ s. Even with modern supercomputer architectures, running for $10^{10}$ MD timesteps to match the DSMC duration is an infeasible task, and more so in the context of concurrent simulations. To overcome this issue, we balance the effects of all time-dependent mechanisms in the MD simulations. Specifically, the interaction of nickel atoms in the balanced region interact with both flow atoms as well as atoms in the bulk region. The former naturally occurs for long simulation duration, and its effect can be investigated to reach an equilibrium. To address the latter, we recognize that classical MD is not able to replicate the radiative heat loss occurring over time. To account for its effect, we must inform the thermostat of the bulk region with analytical formulations. Rewriting the Stefan-Boltzmann law (Eq. \ref{eq:stfboltz}) and considering the heat loss attributed to a temperature change $Q = mc\Delta T$, we obtain:
\begin{align}
    \frac{Q}{\Delta t} &= \sigma \epsilon A T^4 \\
    \frac{mc\Delta T}{\Delta t} &= \sigma \epsilon A T^4 \\
    \Delta T &= \Delta t \frac{\sigma \epsilon A T^4}{\rho Vc} \\
    \Delta T &= \Delta t \frac{\sigma \epsilon T^4}{\rho c d} \ ,
\end{align}
where $\Delta T$ is the resulting temperature change, $\sigma$ is the Stefan-Boltzmann constant, $\epsilon$ denotes the emissivity, $T$ states the temperature of the wire at the beginning of the time step, $\rho$ is the wire's density, $c$ is the specific heat and $d$ is the simulated wire thickness. It is in $\Delta t$ where we introduce the temporal bridging: by setting $\Delta t$ to the duration between consecutive coupling calls (e.g. $\Delta t =\Delta t_{\textrm{DSMC}}$), we emulate radiation effects in the simulation. In practice, this implies that at the beginning of each MD time step, we initialize the wire atoms to a temperature of $T_\textrm{target} = T_\textrm{surf} - \Delta t \frac{\sigma \epsilon T_\textrm{surf}^4}{\rho c d}$. We find that the radiative heat loss and flow particle collisions reach an equilibrium after at least 30,000 MD timesteps. For this simulation, a conservative duration of 50,000 timesteps is therefore chosen to avoid the omission of events which may require longer equilibration times. 

\subsection{Results}
\begin{figure}
\centering
\includegraphics[width=0.7\textwidth]{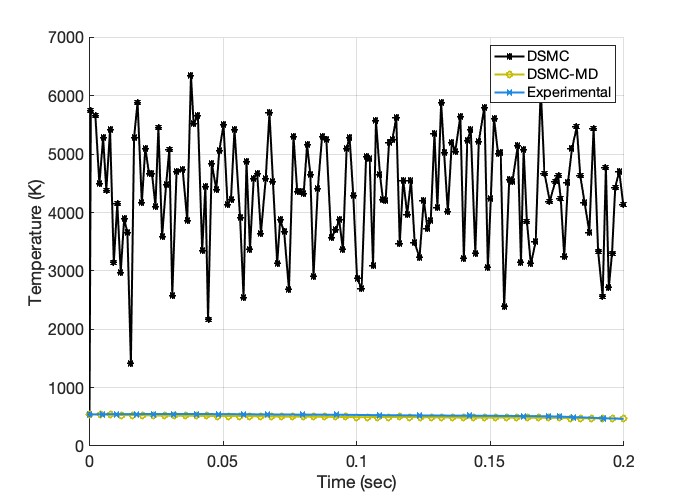}
\caption{Temperature evolution in the hypersonic flow. The DSMC simulation cannot accurately capture the temperature evolution. With the help of a concurrent coupling to MD, it matches experimental data\cite{Wire2013} well.}
\label{fig:wireData}
\end{figure}

The example is an appropriate use case of the coupling mechanism; the wire is very small compared to the flow field, such that microscopic effects may strongly influence the macroscopic flow field. The simulation is run using just DSMC as well as a full DSMC-MD coupling. A comparison to the experimental data is shown in Figure \ref{fig:wireData}.

Figure \ref{fig:wireData} shows the surface temperature resulting from the exposure of the wire to the hypersonic flow for the duration of the experimental data\cite{Wire2013}. The temperature history provided by the DSMC simulation (shown in black) is an order of magnitude larger than the experimental data (blue). In addition, large fluctuations spanning thousands of degrees Kelvin are observed. The mean surface temperature of $\bar{T}_\textrm{DSMC} = 4342.7$ K shows a $730\%$ relative error from the expected mean of $\bar{T}_\textrm{exp} = 523.2$ K. Clearly, the DSMC simulation does not accurately capture the gas-surface interaction, the cause of which is multifaceted. First, the DSMC simulation spans a very large domain, which requires a large value of $f_\mathrm{num}$. Using the only available surface temperature model in SPARTA, the flux of total energy on the wire element is dramatically overestimated with the collision of DSMC particles. Single impacts of DSMC particles simulate the collisions of a very large number of real particles in one instance, which causes the large temperature spikes and fluctuations. Secondly, the large domain size requires the surface element to be underresolved. The surface element is captured by a single grid cell, which further reduces the number of particles subject to impact at any one time. The root of the issue therefore lies in the bridging of length scales in the domain, observed both by the value of $f_\mathrm{num}$ as well as the resolution of the small surface element size compared to the DSMC domain. 

Molecular dynamics provides an excellent alternative. Using a fraction of $f_\mathrm{num}$ atoms, it deterministically simulates the interactions between nickel and flow atoms, which results in a statistically sound temperature evolution. In essence, the MD domain bypasses the inaccurate surface temperature model. By coupling MD to DSMC using the cell-based approach, Figure \ref{fig:wireData} shows that the temperature of the DSMC-MD method (cyan) closely matches the experimental data.

\begin{figure}
\centering
\includegraphics[width=0.65\textwidth]{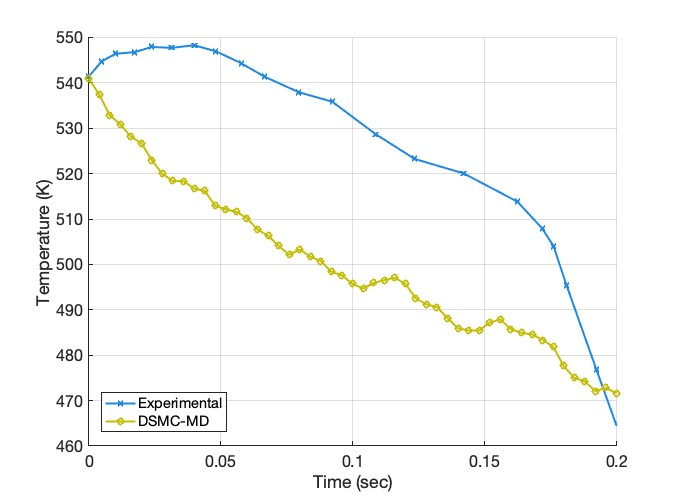}
\caption{A detailed comparison between the coupled DSMC-MD approach and the experimental data. Deviations of up to 7\% are observed.}
\label{fig:wireDataDetailed}
\end{figure}

A more detailed plot of the DSMC-MD coupled data and the experimental data is shown in Figure \ref{fig:wireDataDetailed}. The plot shows the evolution of the surface temperature across the experimental duration for the measured data (blue) \cite{Wire2013} and the fully coupled DSMC-MD simulation (cyan). In both cases, the initial wire temperature of $541$ K is reflected. The experimental data undergoes a brief period of heating, before peaking at a temperature of $548.3$ K. The flow then starts to cool the wire at a steady rate for 0.12 seconds before a sharp drop in temperature is reached. The DSMC-MD coupled data does not show the initial increase in temperature. Instead, the wire appears to experience rapid cooling immediately, with a steady rate achieved after 0.04 seconds. The nature of the discrepancy between the DSMC-MD method and the experiment due to the initial heating will be discussed in more detail later in this section. The mean temperature of the DSMC-MD simulation ($\bar{T}_\textrm{DSMC-MD} = 500.2$ K) generally agrees with the mean temperature of the experimental data ($\bar{T}_\textrm{exp} = 523.2$ K), showing a relative error of $4.4\%$. The maximum discrepancy between the two solutions occurs shortly before the steep cool-off observed in the experiment at $t=0.16$ s. The maximum relative error here is $7\%$. In comparison, the maximum relative error for the standard DSMC simulation showed $1057.8\%$ at $t= 0.038$ s. Therefore, both the mean error and maximum relative error of the coupled DSMC-MD simulation show a significant improvement from the DSMC simulation. 

The advance in accuracy between the standard DSMC and DSMC-MD simulations is especially significant when observing the consequences to the simulation. A high discrepancy in surface temperature will have a direct influence on the density surrounding the surface, as post-collision velocities at the surface will be sampled at dramatically higher magnitudes from the diffuse surface. Since almost all macroscopic quantities in DSMC are higher-order moments of position and velocity, the cascading error throughout all sampled properties would have a dramatic effect. In addition, the gas-surface chemistry is directly impacted by higher surface temperatures. Lastly, the reactivity of particles (described by the Total Collision Energy model\cite{Bird1994}) is driven by the temperature of the particles. With the presented DSMC-MD approach, we provide confidence that the larger flow field as well as microscopic effects, such as gas-surface interactions, are modeled accurately.

The DSMC-MD coupled simulation stills fails to replicate the experimental data. Figure \ref{fig:wireDataDetailed} clearly shows that the evolution of the temperature in the DSMC-MD simulation deviates from the experimental data. Especially at the beginning of the experiment, the initial heating observed in the experimental data is not captured. Additionally, the increase in cooling rate observed in the experiments after 0.16 s is not reflected in the simulation. Failure to replicate these two effects indicates a lack of physical consistency. Given that the change in cooling rate occurs at a similar time ($t=0.04$ s), initial investigations considered the effect of flow density close to the surface: in the experiment, an initial continued increase in density would influence the cooling, and in the simulation the increased collisions of a higher density would explain the shift in cooling rate. However, further investigations into the density field of the simulation showed that an equilibrium is reached after 4 ms - well before the cooling rate change. 

\begin{figure}
\centering
\includegraphics[width=0.65\textwidth]{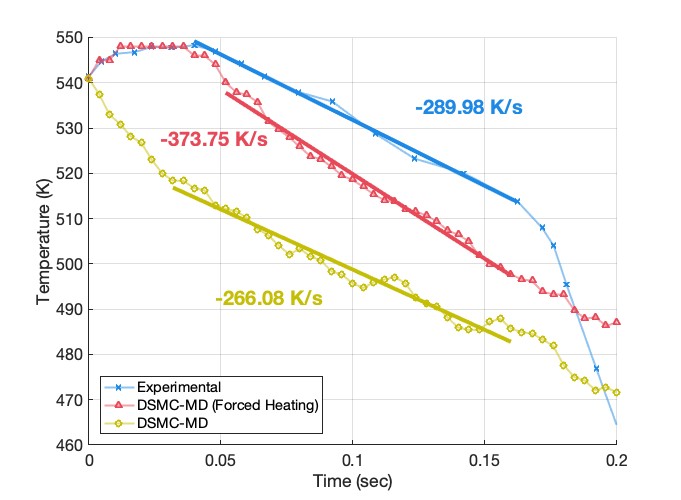}
\caption{Comparison of cooling rates between three approaches. The experimental data is compared to a DSMC-MD simulation as well as a DSMC-MD simulation with forced initial heating.}
\label{fig:wiretempslopes}
\end{figure}

An additional source for the discrepancy could be initial heating effects in the experimental data. To test whether initial heating of the wire in the experiment could also be captured in our DSMC-MD simulation, we applied forced heating to the wire in the initial 0.04 s of the simulation, after which our DSMC-MD coupling was applied. Figure \ref{fig:wiretempslopes} shows the resulting temperature evolution (gray). Again, the DSMC-MD framework fails to capture the exact temperature dynamics. We observe an immediate departure from the experimental data (blue). A steady cooling rate is achieved immediately after, which is maintained throughout the simulation. The rapid temperature drop-off observed in the experiment after $t = 0.16$ s is also not captured. We attribute the increase in cooling rate to the prescribed heating of the wire relative to the flow conditions. It is clearly observed in the DSMC-MD (cyan) data that the flow conditions would cause the wire to cool. The resulting nonequilibrium state drives the cooling to a faster rate than would naturally occur. Nonetheless, we observe a steady cooling rate in all three cases between $0.04 < t < 0.16$ s. Figure \ref{fig:wiretempslopes} depicts the steady region (bold lines) along with its cooling rates. The cooling rates are obtained using linear regression within the steady-state region of each case. Comparing the experimental cooling rate of $-289.98 \frac{\textrm{K}}{\textrm{s}}$ with the DSMC-MD simulation's $-266.08 \frac{\textrm{K}}{\textrm{s}}$ shows a similarity in rates, within an error of $8.2\%$. In contrast, the forced heating shows a much steeper cooling rate of $-373.75 \frac{\textrm{K}}{\textrm{s}}$ which deviates from the expected experimental value by $28.9\%$. We note that due to the short duration of the experiment and simulation, the cooling rate is very sensitive to the measured temperature. We also note that the steady region cooling rate of the standard DSMC simulation (black evolution shown in Figure \ref{fig:wireData}) is, in fact, a heating rate of $+2861.8 \frac{\textrm{K}}{\textrm{s}}$.

Overall, we show qualitative agreement with the experimental data in the steady state region, and significantly improve the standard DSMC formulation through our coupling framework. However, transient effects such as initial heating and rapid increases of cooling at the end of the simulation are not captured. Artificial, forced initial heating in the DSMC-MD simulation to match the initial condition of the steady-state cooling region in the experimental data did not significantly enhance the description. With additional investigation into the verification of our method yielding no further insights, we discuss the experimental data in greater detail.

Widodo \& Buttsworth \cite{Wire2013} show clear evidence that both the initial heating and the rapid cooling physically occur in the wire, and deem the captured gas-surface interaction as correct. To explain the effects, they postulate that macroscopic effects cause the behavior. In their experimental set-up, a diaphragm in the hypersonic test stand is designed to rupture when a desired stagnation pressure is reached. The flow rate can be set such that the stagnation pressure of the nozzle is constant. This is done by setting the pressure in the gas reservoir to equalize the gas expulsion through the hypersonic nozzle, which are known as matched conditions. In their experiments, Widodo \& Buttsworth observed that the average stagnation pressure shows an increase during the test time. They attribute this to overmatched conditions in their experiment. The continued rise of stagnation pressure by over 50\% \cite{Wire2013} is assumed to be a deciding factor for the initial heating of the wire, as it has a strong connection to all thermodynamic variables in the flow. The DSMC simulation set-up does not account for an increase in flow pressure. It sets all experimental values as constant inputs to its initial conditions, and evolves them independently. Therefore, the omission of a continued rise in pressure in the DSMC's initial conditions can be connected to the lack of initial heating. To include this effect, the DSMC simulation would require the definition of variable initial conditions. Such an addition does not pose any changes to the DSMC-MD coupling, as flow conditions are automatically reflected in the MD domain through the lifting operator. It would be a matter of measuring the transient rise in $p$ and $T$ to relay as initial conditions in the DSMC simulation. Its influence on the DSMC-MD coupling is subject to future investigations.

In regard to the steep cooling at the end of the measurement, Widodo \& Buttsworth observed the onset of vortical flows in their tests. Vortical flows cause the flow temperature to drop significantly prior to impeding the surface due to turbulent mixing. These flows are proposed to originate in the vicinity of the piston and move throughout the flow field \cite{Wire2013}. Since the simulation design in DSMC assumes the inflow conditions to be constant, such effects are not propagated throughout the duration of the DSMC simulation. To include these effects, detailed experimental data would be required to replicate the unsteady behavior of the incoming flow. Again, the DSMC-MD coupling mechanism described in this work could inherit all added details fully, and poses no restriction to these macroscopic effects.

In summary, a multiscale method that couples Direct Simulation Monte Carlo and molecular dynamics has a wide range of applications. For large-scale flows that possess complex surface elements, DSMC is able to resolve the flow field in its entirety, while MD captures the gas-surface interactions in great detail. We provided one such example. However, many more examples can be found. During evaporation and condensation processes, both the gas and liquid phases may be captured accurately using a DSMC-MD interface. Novel gas-surface chemistry models can be validated in DSMC using our framework, comparing against chemical kinetics in MD. One-time events such as a foreign particle impact on a surface can be modeled in MD without an explicit model of it in DSMC, e.g., an ice particle impeding a surface in a hypersonic flow. It could also prove useful in fusion reactors for both magnetic and inertial confinement. In magnetic reactors, low collision plasmas can be simulated using DSMC and the complex plasma-blanket interaction is captured using MD. In inertial confinement reactors, the damage and heat transfer done to the chamber by rarefied particles can be modeled using our DSMC-MD framework. Lastly, it could prove helpful in micro- and nanofluidics to accurately capture surface reactions, heat transfer rates, and more. In addition to interesting physical use cases, one may choose to employ a DSMC-MD coupling in cases where a one-to-one mapping is desired without changing $f_\mathrm{num}$. Such cases may occur in events where very few particles are present in a DSMC cell, such that an MD instance would avoid statistical errors. Many DSMC-MD approaches have been implemented for specific problems, and the cell-based coupling approach generalizes the procedure to apply to the wide range of applications.

\section{Conclusion}
This work revisits a longstanding problem in computational physics: the transition between continuum and atomistic scales. Here, we shine new light on employing a particle perspective to bridge the domains. We present the coupling of two particle methods to bridge microscopic insights into macroscopic flow behavior. New lifting and restricting operators serve to translate between scales, and their implementation is explained. By introducing a new ratio between physical to simulated particles in the molecular dynamics domain, we successfully bridge the spatial scale of the two methods. The lifting operator automatically translates domain sizes, atom positions and velocities between domains. The restricting operator obtains the accurate, atomistic information from MD and informs the macroscopic simulation with its findings. We verify the operators through basic conservation laws. Mass is perfectly conserved, and temperature shows fluctuations around its mean within statistical deviations inherent to a stochastic method. An example is provided to showcase the usefulness as well as disadvantages of this approach. We show that the method is unable to capture transient effects by comparing to experimental data, where the initial heating and strong cool-off of the wire is not replicated in the DSMC-MD domain. However, these transient effects may be caused by macroscopic behavior such as an increase in pressure and the onset of vortical flows. In fact, for steady state solutions, we observe that we achieve an improvement of accuracy that is three order of magnitudes better than conventional DSMC simulations. Our method therefore enables the investigation of problems which were unfit for DSMC prior to this.

In general, the integration of molecular dynamics into the functions of the Direct Simulation Monte Carlo method enables a variety of physical problems to be solved more efficiently and accurately, from gas-surface interaction to hydrodynamic instabilities. Its implementation in the popular SPARTA (DSMC)\cite{SPARTA} and LAMMPS (MD)\cite{LAMMPS} software packages enable broad applicability. Further studies may investigate the propagation of transient effects in the macroscopic method to the microscopic instance. The coupling method may be leveraged to include gas composition changes, surface reactions, near-continuum applications, strong shock waves and phase transitions in great detail in a Direct Simulation Monte Carlo simulation. It paves the way for further microscopic insights into macroscopic effects.

\section*{Acknowledgments}
T. Linke thanks the Presidential Fellowship at LLNL for funding this work under Task No. PECASE002. This work was performed under the auspices of the U.S. Department of Energy by Lawrence Livermore National Laboratory under Contract No. DE-AC52-07NA27344.

\section*{Supplemental Materials}
A video highlighting the coupling process can be found at \newline \url{https://youtu.be/-KR9Ebh1xWY}. The method is made available through the open-source distribution of SPARTA under the command \textit{fix lammps}. More information can be found at \url{https://github.com/sparta/sparta/pull/556}.

\section*{CRediT authorship contribution statement}
\textbf{Tim Linke}: Conceptualization, Methodology, Software, Formal analysis, Writing- Original draft. \textbf{Dane Sterbentz}: Conceptualization, Methodology, Formal analysis, Writing- Review \& Editing. \textbf{Niels Grønbech-Jensen}: Methodology, Formal analysis, Writing- Review \& Editing. \textbf{Jean-Pierre Delplanque}: Conceptualization, Supervision, Formal analysis, Writing- Review \& Editing, Funding acquisition. \textbf{Jonathan Belof}: Conceptualization, Supervision, Project administration, Funding acquisition.

\end{document}